\documentclass[12pt,preprint,showpacs,prd,nofootinbib]{revtex4}
\usepackage{graphicx}
\usepackage{psfrag}
\usepackage{amssymb,amsfonts,amsmath}

\newcommand{\dd}{\mathrm{d}}
\newcommand{\ee}{\mathrm{e}}
\newcommand{\ii}{\mathrm{i}}

\begin{document}

\title{Collisions of rigidly rotating disks of dust in General
Relativity}

\author{J\"org Hennig}
\email[e-mail: ]{J.Hennig@tpi.uni-jena.de}

\author{Gernot Neugebauer}

\affiliation{Theoretisch-Physikalisches Institut, Friedrich-Schiller-Universit\"at Jena, Max-Wien-Platz 1, D-07743 Jena, Germany}

\date{\today}

\begin{abstract}
We discuss inelastic collisions of two rotating disks by using the
conservation laws 
for baryonic mass and angular momentum. In particular, we formulate
conditions for the formation of a new disk after the collision and
calculate the total energy loss to obtain upper limits for the emitted
gravitational energy. 

\end{abstract}

\pacs{04.20.Jb, 04.30.Db, 04.25.Nx, 95.30.Sf}

\maketitle

\section{Introduction}

Disk-like matter configurations play an
important role in astrophysics (e.g. as models for galaxies, accretion
disks or intermediate
phases in the merger process of two neutron stars).
The simplest models for such configurations are disks of dust.
From a mathematical point of view they are
solutions to boundary value
problems of the Einstein equations.
Explicit solutions in terms of standard functions or integrals are known
for \emph{rigidly rotating disks of dust} \cite{Neugebauer1, Neugebauer2,
Neugebauer3, Bardeen1, Bardeen2}  (the only known solutions for isolated
rigidly rotating bodies)
and \emph{counter-rotating disks of dust}, consisting of clockwise
and counter-clockwise rotating dust particles (see e.g. \cite{Morgan}, \cite{Bicak} or \cite{Meinhardt}).

In this paper we want to study inelastic collisions of rotating
disks. (Due to the gravitational radiation there exist no \emph{elastic}
collisions in General Relativity.)
A rigorous mathematical description of such merging processes is
extremely difficult, for the solution of the corresponding initial boundary
value problem requires extensive numerical investigations. We adopt a different
method and perform a \lq\lq thermodynamic\rq\rq\ analysis. In this way we
forgo the detailed analysis in favour of a simpler description leading
to a \lq\lq rough\rq\rq\ picture of the merging processes.
A classical example for this procedure was given by
Hawking and Ellis who 
discussed the efficiency of the collision and
coalescence of two black holes, cf. \cite{Hawking}.
By using the area theorem for black holes they
obtained, for spherically symmetric black holes,
an upper limit for the efficiency of conversion of
mass into gravitational radiation of $1-1/\sqrt{2}\approx 29.3\%$.
One of our results will be a similar limit for the
coalescence of two disks as an example for \lq\lq normal matter\rq\rq\
collisions. Considerations like these are typical for thermodynamics, in
which initial and final equilibrium states are linked by conserved
quantities bridging the intermediate non-equilibrium states of the
system. Interestingly, a Gibbs equation for the thermodynamical potential
\emph{energy (-mass)} $M$ as a function of baryonic mass $M_0$ and angular
momentum $J$ can be formulated even in the case of rotating dust matter,
see Eq. \eqref{1c0}.

\begin{figure}
\resizebox{0.9\textwidth}{!}{\includegraphics{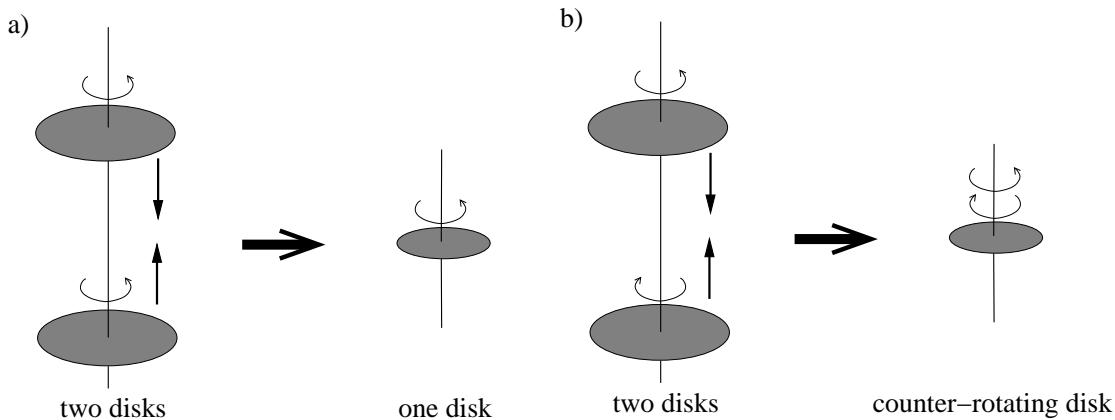}}
\caption{Illustration of the two collision scenarios}
\label{figure1}
\end{figure}

In particular,
we study the \lq\lq head on\rq\rq\ collision of two aligned rigidly rotating
disks of dust with parallel [scenario (a)] or antiparallel [scenario
(b)] angular momenta, cf. Fig.~\ref{figure1}.
Rigid rotation is a universal limit for
rotating disks of dust.
Any amount of friction between the rings comprising the disk
of dust will lead to  an
equilibrium state with constant angular velocity $\Omega$ after a
sufficiently long time.  As a consequence, \emph{rigidly} rotating disks
are characterized
by an extremum in the binding energy, compared to \emph{differentially}
rotating disks with the same baryonic mass and angular momentum,
see  appendix \ref{Extremal}.
We assume that the initial distance between the disks is large enough to
keep the initial gravitational interaction very small.

As already
mentioned, the dynamics of
the collision process is outside the scope of our
considerations. However, we know that the total baryonic mass $M_0$ and the
total angular momentum $J$ are conserved. Finally, due to the outgoing
gravitational radiation and possible dissipative processes, the
collision ends in a stationary (and axisymmetric) configuration with the
total baryonic mass $M_0$ and the total angular momentum $J$.

In this paper we confine ourselves to two problems:
the formation of a rigidly
rotating disk (RR disk)
from two rigidly rotating initial disks and, as a second
example, the formation of the rigidly counter-rotating disk 
described by Bardeen, and Morgan and Morgan \cite{Morgan} (RCR disk)  from two
rigidly rotating initial disks with opposite angular momenta.
We will discuss questions like these:
For which parameter values of the initial disks can the
collision lead to a rigidly rotating or rigidly counter-rotating
disk at all?  Which domain of the
$M_0$-$J$-parameter space can be
reached by such processes?
As we will see in section \ref{crdisk}, 
the relative
binding energy $E_\textrm{b}$ of the rigidly counter-rotating disks
takes positive as well as  negative
values. Therefore, it is an interesting question if the
formation of RCR disks with
a negative binding energy
is a possible result of
collision processes.
Finally, we calculate upper
limits for the energy loss due to gravitational radiation in such collision
processes.

The mathematical analysis of these problems requires the examination of
the \lq\lq thermodynamics\rq\rq\ of
\emph{rigidly rotating disks}  and
\emph{rigidly counter-rotating disks} and the discussion of
the equations of state for
mass and angular momentum. This is done
in section \ref{thermodynamics}.   
In section \ref{conservation} we discuss limits for the formation of
disks after the collision.
The conservation equations can be used to calculate the parameters of the
merged disks in terms of elliptic functions. This analysis and the
resulting plots can be found in section \ref{properties}.
The energy loss, i.e. the efficiency of the two
scenarios is calculated in section
\ref{efficiency}.

The metric coefficients of the rigidly rotating disk of dust solution
are given in terms of ultraelliptic theta functions which reduce
to elliptic functions along the axis of symmetry. Since the multipole
moments for energy (-mass) $M$ and angular momentum $J$, as the central
quantities of our thermodynamic considerations, can be read off from the
axis values of the metric, our analysis has to make extensive use of
elliptic functions. To avoid lengthy calculations in the main body of
the text, we relegate the analytic expressions to the appendix and present the
results in graphical form in the main text.

For the sake of simplicity, we restrict ourselves to
identical initial disks (i.e. disks of equal baryonic mass and equal
absolute values for the angular momenta). The generalization to
different disks is a straightforward procedure.

\section{\lq\lq Thermodynamics\rq\rq\ of disk models}\label{thermodynamics}
\subsection{Disk of dust solutions}
\subsubsection{Rigidly rotating disks of dust}\label{disk}
This section is devoted to a thermodynamic description of the
\lq\lq ingredients\rq\rq\ of the collision processes: rigidly rotating
disks (RR disks) and rigidly counter-rotating
disks (RCR disks).

The free boundary value problem for the relativistic rigidly
rotating disk of dust was approximately discussed by Bardeen and Wagoner
\cite{Bardeen1,Bardeen2} and
analytically solved in terms of ultraelliptic theta functions
by Neugebauer and Meinel
\cite{Neugebauer1,Neugebauer2,Neugebauer3} using the Inverse Scattering
Method. For a discussion of the
physical properties see \cite{Neugebauer4}.
The solution is stationary (Killing vector: $\xi^i$) and axisymmetric
(Killing vector: $\eta^i$). Its line element together with the Killing
vectors can therefore be written in the Weyl-Lewis-Papapetrou standard form
\begin{equation}\label{1}
\dd s^2=\ee^{-2U}[\ee^{2k}(\dd
\rho^2+\dd\zeta^2)+\rho^2\dd\varphi^2]-\ee^{2U}(\dd t+a\dd\varphi)^2,
\quad \xi^i=\delta^i_t,\quad \eta^i=\delta^i_\varphi,
\end{equation}
where $U$, $k$ and $a$ are functions of  $\rho$ and
$\zeta$ alone  and $\delta^i_k$ is the four-dimensional Kronecker symbol.
Note that we use the normalized units where $c=1$ for the speed of light and
$G=1$ for Newton's gravitational constant. 
The solution can be written in terms of the complex Ernst
potential $f(\rho,\zeta)=\ee^{2U(\rho,\zeta)}+\ii b(\rho,\zeta)$, where
the imaginary part is related to $a$ by
$a_{,\rho}=\rho\ee^{-4U}b_{,\zeta}$ and
$a_{,\zeta}=-\rho\ee^{-4U}b_{,\rho}$. In this formulation,
the vacuum Einstein equations
are equivalent to the Ernst equation
\begin{equation}\label{1.0}
(\Re f)(f_{,\rho\rho}+f_{,\zeta\zeta}+\frac{1}{\rho}f_{,\rho})=f_{,\rho}^2+f_{,\zeta}^2.
\end{equation}
($k$ can be calculated via a path integral from the Ernst potential $f$.)

The matter of the disk of dust is described by the energy-momentum tensor
\begin{equation}\label{1.1}
T^{ij}=\varepsilon u^i u^j,\quad \varepsilon=\sigma(\rho)
\delta(\zeta),
\end{equation}
where $\varepsilon$, $\sigma$ and
$u^i$ are the mass density, the surface mass density
($\sigma(\rho)=0$ if $\rho>\rho_0$, $\rho_0$ being the
coordinate radius of the disk) and the
four-velocity of the dust particles, respectively.
For rigidly rotating bodies,
the four-velocity is a linear combination of the two killing
vectors,
\begin{equation}\label{1.2}
u^i=\ee^{-V_0}(\xi^i+\Omega\eta^i),
\end{equation}
where $\Omega$ is the constant angular velocity of the body.
Here, as a
consequence of the geodesic motion of the dust particles,
the coefficient $\ee^{-V_0}$ turns out to be a constant too,
\begin{equation}\label{1.2a}
V_0=\textrm{constant}.
\end{equation}

The RR disk solution depends on two parameters.
As an example, one may choose
the coefficients $\ee^{-V_0}$ and $\Omega\ee^{-V_0}$
of the linear combination \eqref{1.2} or, alternatively
the coordinate radius $\rho_0$ of the disk
and a centrifugal parameter
$\mu=2\Omega^2\rho_0^2\ee^{-2V_0}$
($\mu\to 0$ turns out to be the
Newtonian limit and $\mu\to 4.62966\dots$ the ultrarelativistic limit,
where the disk approaches the extreme Kerr black hole,
cf. \cite{Neugebauer2} and \cite{Neugebauer4} for these and further properties).

The baryonic mass $M_0$, the gravitational (ADM) mass $M$ and the angular
momentum $J$ of the disk are given by
\begin{equation}\label{1a}
M_0=\int\limits_\Sigma\varepsilon\sqrt{-g}u^t\dd^3 x,
\end{equation}
\begin{equation}\label{1b}
M=2\int\limits_\Sigma\left(T_{ij}-\frac{1}{2}Tg_{ij}\right)n^i\xi^j\dd V,
\end{equation}
\begin{equation}\label{1c}
J=-\int\limits_\Sigma T_{ij}n^i \eta^j \dd V,
\end{equation}
with $T_{ij}$ as in \eqref{1.1}.
$\Sigma$ is the spacelike hypersurface $t=\mathrm{constant}$ with
the unit future-pointing normal vector $n^i$.

\subsubsection{Rigidly counter-rotating disks of dust}\label{crdisk}
An interesting example of a counter-rotating disk is the RCR disk
by Bardeen, Morgan and Morgan \cite{Morgan}, consisting of a clockwise
and a counter-clockwise rotating
component of dust. All mass elements move along geodesic lines and the
two components have constant angular velocities with opposite signs.
Since the net angular momentum of the disk vanishes, its metric can be
written in the \emph{static} Weyl-Lewis-Papapetrou form [$a=0$ in \eqref{1}],
\begin{equation}\label{5}
\dd s^2=\ee^{-2U}[\ee^{2k}(\dd
\rho^2+\dd\zeta^2)+\rho^2\dd\varphi^2]-\ee^{2U}\dd t^2,
\quad \xi^i=\delta^i_t,\quad \eta^i=\delta^i_\varphi,
\quad \xi^i\eta_i=0.
\end{equation}
The metric functions $U$ and $k$ are standard integrals,
see appendix \ref{AppendixB}.
Here the energy-momentum tensor is the superposition of two expressions
\eqref{1.1},
\begin{equation}\label{16.0}
T^{ij}=\frac{1}{2}\varepsilon(u^i u^j+v^i
v^j),\quad u^i=\ee^{-V_0}(\xi^i+\Omega\eta^i),\quad
v^i=\ee^{-V_0}(\xi^i-\Omega\eta^i),
\end{equation}
where $u^i$ and $v^i$ denote the four-velocities of the
counter-clockwise and clockwise moving dust particles.
Just as for the case of the RR disk, $V_0$ turns out to be a
constant as a consequence of the constant angular
velocity $\Omega$ and the geodesic motion of the dust particles,
and the solution is again governed by two parameters.
As with the RR disks, these could be chosen to be
the coordinate radius $\rho_0$
of the disk and the 
centrifugal parameter $\mu=2\Omega^2\rho_0^2\ee^{-2V_0}$. However, it turns out
that, instead of $\mu$, the parameter $b=\Omega\rho_0\ee^{-2V_0}$
simplifies the discussions. ($b\to 0$ is the
Newtonian limit and $b\to \infty$ the ultrarelativistic limit.)

The baryonic mass $M_0$ and the gravitational mass $M$ of the RCR disk
and the angular momenta $\pm J$ of the counter-clockwise and clockwise
rotating part (the resulting angular momentum vanishes) can again
be calculated from the
Eqs. \eqref{1a}-\eqref{1c} (where in the formula for $J$ only the
energy-momentum tensor of the counter-clockwise rotating dust component
is used). For the calculation we refer to appendix \ref{AppendixA1}.

\subsection{Equilibrium and stability}\label{EquilibriumStability}
Equilibrium configurations can be described with the aid of variational
principles (cf. \cite{Hartle,NeugebauerTD}). For disks of dust we may
consider the thermodynamic potential
\begin{equation}\label{E1}
E:=\int\limits_{t=t_0}\left(\frac{R}{8\pi}+\varepsilon\right)\sqrt{-g}\,\dd^3 x
+n\Omega J+\frac{M}{2},
\end{equation}
where $R$ is the Ricci scalar and $n$ indicates the number of dust
components 
($n=1$ for RR disks and $n=2$ for RCR disks). The variation of $E$
leads to
\begin{equation}\label{E2}
\delta E=-\frac{1}{8\pi}\int\left(R^{ij}-\frac{R}{2}g^{ij}+8\pi
T^{ij}\right)
\sqrt{-g}\delta g_{ij}\,\dd^3 x+\ee^{V_0}\delta M_0+n\Omega\delta J,
\end{equation}
with $T^{ij}$ from \eqref{1.1} or \eqref{16.0} and $M_0$, $M$ and $J$
from \eqref{1a}-\eqref{1c}. Obviously,
for fixed baryonic mass
$M_0$ and fixed angular momentum $J$, i.e. $\delta M_0=0$ and $\delta
J=0$,  the condition $\delta E\big|_{M_0,J}=0$ leads to
the Einsteinian field equations. On the other hand, for solutions to the
field equations it turns out that $E$ is equal to the gravitational mass
$M$. Hence, one obtains
\begin{equation}\label{E3}
\delta M=\ee^{V_0}\delta M_0+n\Omega\delta J.
\end{equation}

To illustrate the meaning of the potential $E$ we should mention
that the Newtonian limit of Eq. \eqref{E1} is given by
\begin{equation}\label{E4}
E=M_0+\frac{1}{2}\int\rho^\textrm{N}U^\textrm{N}\,\dd^3x+n\frac{J^2}{2I},
\end{equation}
where $\rho^\textrm{N}$, $U^\textrm{N}$ and $I$
are the Newtonian mass density, the Newtonian gravitational potential and
the moment of inertia of the disk, respectively.
In this limit, $E$ is the sum of
the rest energy $M_0c^2$ ($c=1$),
the potential energy and the rotational energy.
The function $E$ in
Eq. \eqref{E4} is precisely the quantity that was used by Katz in
\cite{Katz} to study equilibrium and stability of Maclaurin
and Jacobi ellipsoids in Newtonian theory. In 
subsection \ref{AppRCR}
we will use the relativistic generalisation \eqref{E1} of $E$
to investigate the stability of counter-rotating (RCR) disks of dust.

\subsection{Applications}

\subsubsection{Rigidly rotating disks of dust}\label{AppRR}
The extensive quantities $M_0$, $M$ and $J$ are not independent of each
other. Due to Eq.~\eqref{E3} (with $n=1$) they have to satisfy the Gibbs formula
(see also \cite{Bardeen2}, \cite{NeugebauerTD} or \cite{Demianski})
\begin{equation}\label{1c0}
\dd M=\ee^{V_0}\dd M_0+\Omega\dd J,
\end{equation}
where $M=M(M_0,J)$ is a potential in $M_0$, $J$.
Other potentials can be obtained via Legendre transformations. As an
example we may use, as a consequence of \eqref{1a}-\eqref{1c},
the parameter relation
\begin{equation}\label{1c1}
M=\ee^{V_0}M_0+2\Omega J
\end{equation}
to eliminate $M$ in \eqref{1c0}. We arrive at
\begin{equation}\label{1d}
\dd (\Omega J)=-J\dd\Omega-\ee^{V_0} M_0\dd V_0,
\end{equation}
and can now use the potential $\Omega J$ to calculate
$J$ and $M$ as functions of the pair $\Omega$ and $V_0$ or,
alternatively,  $\Omega$ and $\mu$.
To get explicit expressions for these \lq\lq equations of state\rq\rq\
we make use of the disk of dust solution
\cite{Neugebauer1,Neugebauer2,Neugebauer3} to obtain
\begin{equation}\label{A0}
\Omega
J=\frac{\ee^{V_0}}{4\Omega}\int\limits_0^\mu
 \ee^{-V_0(x)} b_0'(x)\,\dd x -\frac{1}{4\Omega}b_0(\mu),
\end{equation}
where $b_0$ is the imaginary part of the Ernst potential in the center
of the disk, $b_0=b(\rho=0,\zeta=0^+)$,
and, as a consequence of \eqref{1d},
\begin{equation}\label{1e}
M_0(\Omega,\mu)=-\ee^{-V_0}\left.\frac{\partial (\Omega J)}{\partial
V_0}\right|_\Omega = -\frac{1}{4\Omega}\int\limits_0^\mu
 \ee^{-V_0(x)} b_0'(x)\,\dd x,
\end{equation}
\begin{equation}\label{1f}
J(\Omega,\mu)=-\left.\frac{\partial (\Omega J)}{\partial
\Omega}\right|_{V_0}=\frac{\ee^{V_0}}{4\Omega^2}\int\limits_0^\mu
 \ee^{-V_0(x)} b_0'(x)\,\dd x -\frac{1}{4\Omega^2}b_0(\mu).
\end{equation}
The parameter relation \eqref{1c1} takes the form
\begin{equation}\label{1g}
M=\ee^{V_0}M_0+2\Omega J=-\ee^{V_0}M_0-\frac{1}{2\Omega}b_0(\mu).
\end{equation}
For explicit expressions for the metric coefficients $b_0(\mu)$,
$V_0(\mu)$, the masses $M_0$
and $M$ and the angular momentum $J$ in terms of elliptic functions
see appendix~\ref{AppendixA}.

There is an interesting scaling behaviour of the disk parameters.
For example $\Omega\rho_0$, $\Omega M$, $\Omega M_0$, $\Omega^2 J$, $M/M_0$,
$M^2/J$ and $M_0^2/J$ depend only on the centrifugal parameter $\mu$ but
not on a second parameter, cf. \eqref{1e}-\eqref{1g} and
appendix \ref{AppendixA}.

\begin{figure}
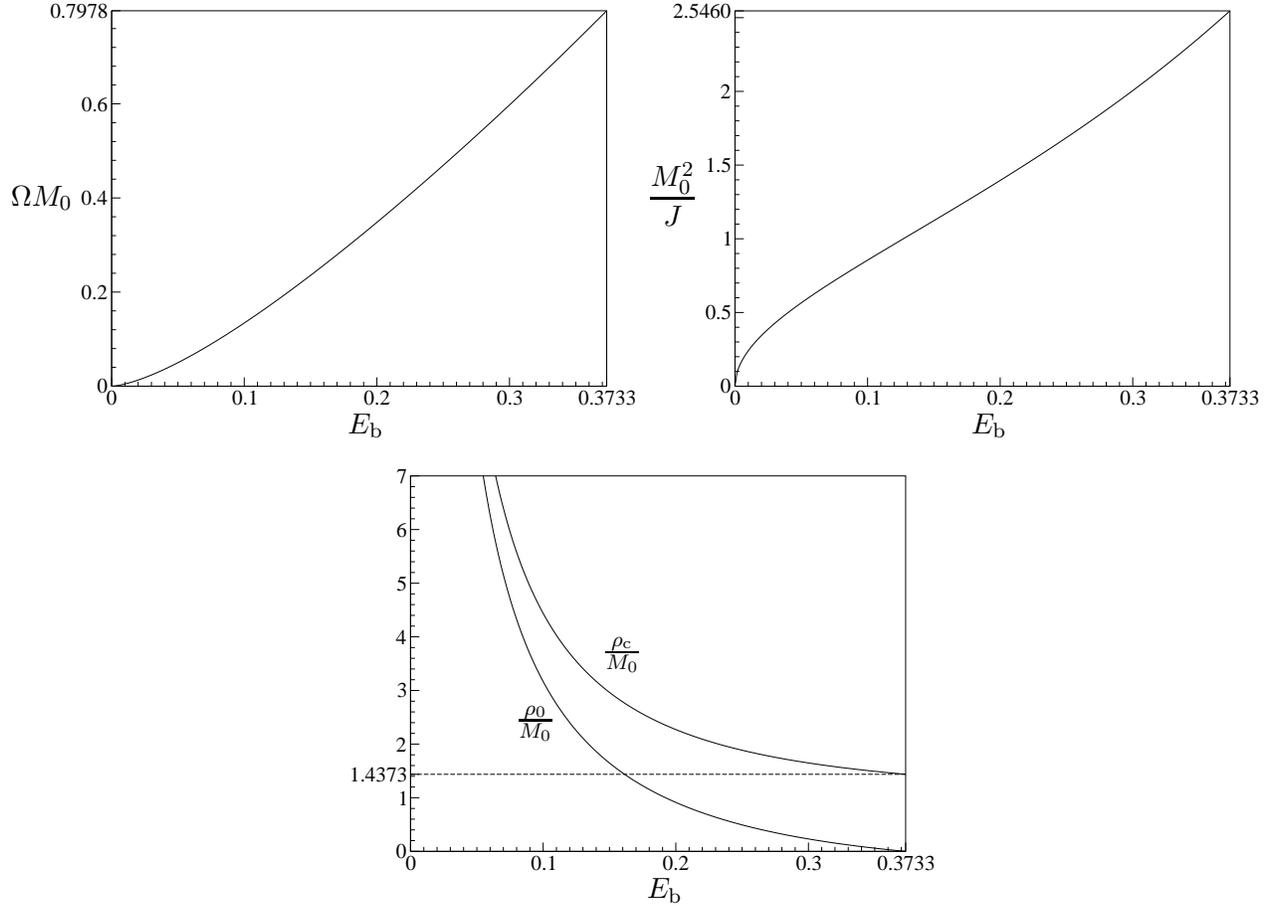

\psfrag{A1}{$\!\!\!\Omega M_0$}
\psfrag{A2}{$\displaystyle \frac{M_0^2}{J}$}
\psfrag{B1}{$\frac{\rho_\textrm{c}}{M_0}$}
\psfrag{B2}{$\frac{\rho_0}{M_0}$}
\psfrag{Eb}{$E_\textrm{b}$}
\includegraphics[scale=0.28]{fig2a.eps}
\includegraphics[scale=0.28]{fig2b.eps}\\[2.5ex]
\includegraphics[scale=0.28]{fig2c.eps}
\caption{$\Omega M_0$, $M_0^2/J$ and the radii $\rho_0$ and
$\rho_\textrm{c}$ as functions of the relative binding
energy $E_\textrm{b}=1-M/M_0$ for the rigidly rotating disk of dust.}
\label{Staubscheibe}
\end{figure}
For an illustration of the 
equations of state Fig. \ref{Staubscheibe} shows
relations between $\Omega M_0$, $M_0^2/J$ and the relative binding
energy $E_\textrm{b}=1-M/M_0$.
The figure also displays the coordinate radius $\rho_0$ and the \lq\lq
circumferential radius\rq\rq\ $\rho_\textrm{c}$ (defined
as $\rho_\textrm{c}=\frac{1}{2\pi}\int\dd s|_{\rho=\rho_0, t=t_0, \zeta=0}
=\sqrt{g_{\varphi\varphi}}|_{\rho=\rho_0, t=t_0, \zeta=0}$).
Thereby $E_\textrm{b}\to 0$
and $E_\textrm{b}\to 0.3733\dots$ are the Newtonian and the
ultrarelativistic limit, respectively. The picture demonstrates the
\lq\lq parametric collapse\rq\rq\ of a disk towards the black hole limit
\cite{NeugebauerAnn}:
Consider a disk with a fixed number of baryons (i.e. fixed $M_0$)
occupying states with decreasing energy $M$. Then, the angular velocity
increases and the angular momentum decreases
while the disk shrinks ($\rho_0\to 0$,
while the \lq\lq true\rq\rq\ radius remains strictly positive,
$\rho_\textrm{c}\to 1.4372\dots\mbox{}\!\cdot M_0$).
In the limit
$E_\textrm{b}=0.3733\dots$, $M_0^2/J=2.5460\dots$ one obtains a ratio 
$M^2/J=M_0^2/J\cdot(1-E_\textrm{b})^2=1$ corresponding to the extreme Kerr
black hole \cite{NeugebauerAnn}.

\subsubsection{Rigidly counter-rotating disks of dust}\label{AppRCR}
As in the case of RR disks we can formulate a Gibbs relation for RCR
disks, too. From Eq.~\eqref{E3} (with $n=2$) we obtain
\begin{equation}\label{16a0}
\dd M=\ee^{V_0}\dd M_0+2\Omega\dd J.
\end{equation}
A Legendre transformation leads to the potential $\Omega J$ satisfying
the equation
\begin{equation}\label{16a1}
\dd (\Omega J)=-J\dd\Omega-\frac{1}{2}\ee^{V_0} M_0\dd V_0,
\end{equation}
From the analytic solution we find
\begin{equation}\label{16a}
\Omega
J(\Omega,V_0)=\frac{1}{8\pi\Omega}\int\limits_0^{\sqrt{\ee^{-4V_0}-1}}\left(\frac{1}{\sqrt{1+t^2}}-\ee^{2V_0}\right)\frac{t\arctan
t}{1+t^2}\,\dd t.
\end{equation}
This potential can be used to calculate the baryonic mass $M_0$ and the angular
momentum $J$ via
\begin{equation}\label{16b}
M_0=-2\ee^{-V_0}\left.\frac{\partial (\Omega J)}{\partial
V_0}\right|_\Omega,\quad
J=-\left.\frac{\partial (\Omega J)}{\partial
\Omega}\right|_{V_0},\quad
M=\ee^{V_0}M_0+4\Omega J,
\end{equation}
where $V_0$ and $\Omega$ are related to the parameters $\rho_0$ and
$b$ by the equations
\begin{equation}\label{16b1}
\ee^{-4V_0}=1+4b^2,\quad \Omega\rho_0=b/\sqrt{1+4b^2}.
\end{equation}

The RCR disks show the same scaling behaviour of the disk parameters
as the RR disks, e.g.
$\Omega\rho_0$, $\Omega M$, $\Omega M_0$, $\Omega^2 J$, $M/M_0$,
$M^2/J$ and $M_0^2/J$ depend only on the parameter $b$ and
not on the coordinate radius $\rho_0$.

\begin{figure}
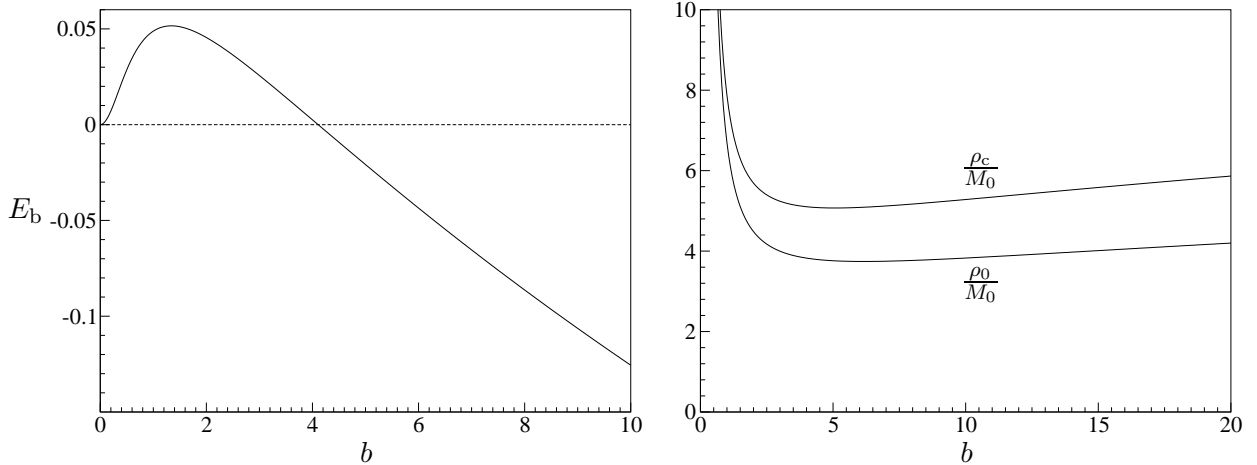

\psfrag{b}{$b$}
\psfrag{B1}{$\frac{\rho_\textrm{c}}{M_0}$}
\psfrag{B2}{$\frac{\rho_0}{M_0}$}
\psfrag{Eb}{$E_\textrm{b}$}
\includegraphics[scale=0.30]{fig3a.eps}\quad
\includegraphics[scale=0.30]{fig3b.eps}
\caption{Left picture: The relative binding energy
$E_\textrm{b}=1-M/M_0$ of the RCR disk as a function of the centrifugal
parameter $b$. $E_\textrm{b}$
reaches negative values in the relativistic region (large $b$).
Right picture: The coordinate radius $\rho_0$ and the \lq\lq
circumferential radius\rq\rq\ $\rho_\textrm{c}$ divided by the baryonic
mass $M_0$ as functions of the centrifugal parameter $b$.} 
\label{Bindungsenergie}
\end{figure}

An interesting feature of the RCR disks is the strange behaviour of the
binding energy, which is negative for $b>4.1074\dots$
as shown in Fig.~\ref{Bindungsenergie}.
Interestingly, there are several physical arguments against the
formation of RCR disks in this region. In fact, the following
application of the stability analysis of Sec. \ref{EquilibriumStability}
will show a transition from stability to instability at
$b=1.3393\dots$,
even before the binding energy $E_\textrm{b}$ becomes negative.
Moreover, using 
a general argument
resulting from the baryonic mass conservation we will show in the next
section that RCR disks with negative binding energy
cannot form by collision.

Fig.~\ref{Bindungsenergie} also
shows $\rho_0/M_0$ and $\rho_\textrm{c}/M_0$
as functions of $b$ (right picture), with $\rho_\textrm{c}$ defined as
in subsection \ref{AppRR}. For
increasing $b$ the radii first decrease, reach a minimum and increase
again. There is no \lq\lq parametric collapse\rq\rq\ of
the RCR disks towards a black hole.

In order to understand
the strange behaviour of the binding
energy we now apply the stability analysis of
Sec. \ref{EquilibriumStability} to RCR disks. Calculating the Ricci
scalar $R$ of the
metric \eqref{5} we obtain from \eqref{E1}
\begin{equation}\label{E5}
E=\int \left[\frac{1}{8\pi}(\nabla U)^2+\varepsilon\sqrt{-g}\right]
\dd\rho\dd\varphi\dd\zeta
+2\Omega J.
\end{equation}
Note that the term $M/2$ in  \eqref{E1} is compensated by a surface
term resulting from the integration over $R$.
The variation $\delta E\big|_{M_0,J}=0$ leads to the equations
\begin{equation}\label{E6}
\bigtriangleup U=4\pi S(\rho)\delta(\zeta),\quad
S(\rho)\delta(\zeta):=-\ee^{2k-2U}(T^t_t-T^\varphi_\varphi)
=\varepsilon \ee^{2k-2V}(1+\Omega^2\rho^2\ee^{-4U}),
\end{equation}
\begin{equation}\label{E5a}
M_0=\int\varepsilon\ee^{-V}\sqrt{-g}\,\dd\rho\dd\varphi\dd\zeta=
\textrm{constant},\quad
J=\int\frac{\varepsilon}{2}\ee^{-V}\eta_i u^i
\sqrt{-g}\,\dd\rho\dd\varphi\dd\zeta=
\textrm{constant}.
\end{equation}
Using the relation \eqref{E6} and the equation
$\ee^{2V}=\ee^{2U}(1-\Omega^2\rho^2\ee^{-4U})$ as a consequence of
$u^iu_i=v^iv_i=-1$,  \eqref{E5} can be
rewritten as
\begin{equation}\label{E7}
E=2\pi\int
\left[-\frac{1}{2}S U+\frac{\ee^{4U}-\Omega^2\rho^2}
{\ee^{4U}+\Omega^2\rho^2}S   \right]
\dd\rho+2\Omega J.
\end{equation}
In the same way we calculate the baryonic mass $M_0$ and the angular
momentum $J$ as defined in \eqref{E5a},
\begin{equation}\label{E6b}
M_0=2\pi\int\limits_0^{\rho_0}
\frac{\sqrt{\ee^{4U}-\Omega^2\rho^2}}{\ee^{4U}+\Omega^2\rho^2}
S(\rho)\ee^U\rho\dd\rho,
\end{equation}
\begin{equation}\label{E6c}
J=\pi\Omega\int\limits_0^{\rho_0}\frac{S(\rho)}{\ee^{4U}+\Omega^2\rho^2}
\rho^3\dd\rho.
\end{equation}

Now we consider the two-parametric family of functions
\begin{equation}\label{E8}
S(\rho;\rho_0,b)=\frac{b}{\pi^2\rho_0}\frac{1}
{\sqrt{1+4b^2\frac{\rho^2}{\rho_0^2}}}
\arctan\frac{2b\sqrt{1-\frac{\rho^2}{\rho_0^2}}}
{\sqrt{1+4b^2\frac{\rho^2}{\rho_0^2}}}
\end{equation}
and the corresponding disk potential [as a solution of \eqref{E6}]
\begin{equation}\label{E9}
U(\rho;\rho_0,b)=\frac{1}{2}\ln\frac{1+{\sqrt{1+4b^2\frac{\rho^2}{\rho_0^2}}}}
{2\sqrt{1+4b^2}},
\end{equation}
where $b$ and $\rho_0$ are arbitrary constants. Obviously, $E$ depends
on the parameters $b$,  $\rho_0$ and $\Omega$, $E=E(b,\rho_0,\Omega)$.
It should be emphasized that $S(\rho;\rho_0,b)$ and $U(\rho;\rho_0,b)$
define a family of trial functions which do \emph{not} satisfy the 
equation \eqref{E5a}.
Only if
$b$, $\rho_0$ and $\Omega$ are connected by the relation \eqref{16b1}
we arrive at the RCR disk solution. 
To calculate the extremum of $E$ for fixed values of $M_0$, $J$ ($\delta
E\big|_{M_0,J}=0$) we replace $\rho_0$ and $\Omega$ via \eqref{E6b},
\eqref{E6c} by $M_0$ and $J$. With the explicit expressions \eqref{E8} and
\eqref{E9}, $M_0$, $J$ and $E$ take the form
\begin{equation}\label{E9a}
M_0=\rho_0 g_1(b,\Omega\rho_0),\quad
J=\Omega\rho_0^3 g_2(b,\Omega\rho_0),\quad
E=\rho_0g_3(b,\Omega\rho_0)+2\Omega J
\end{equation}
with functions $g_1$, $g_2$ and $g_3$ expressible 
in terms of integrals resulting
from \eqref{E7}-\eqref{E6c}. The combination
\begin{equation}\label{E9b}
s:=M_0^2/J=\frac{g_1(b,\Omega\rho_0)}{\Omega\rho_0 g_2(b,\Omega\rho_0)}
\end{equation}
allows to express $\Omega\rho_0$ in terms of $b$ and $s$,
$\Omega\rho_0=g_4(b;s)$. 
Hence, we finally obtain $E$ in
the form
\begin{equation}\label{E10}
E=M_0\frac{g_3[b,g_4(b;s)]+2g_4^2(b;s)g_2[b,g_4(b;s)]}{g_1[b,g_4(b;s)]}
=:-M_0 \tilde E(b;s), 
\end{equation}
where the dependence of $\tilde E$
on $b$ and $s$ is given implicitly by some integral
relations.
The minima of $E$ for fixed $M_0$ and $J$ are the maxima of $\tilde E$
for fixed $s$ and vice versa.

\begin{figure}
\psfrag{b}{$b$}
\psfrag{K(s)}{$K(s)$}
\psfrag{s}{$s$}
\psfrag{A}{A}
\psfrag{unstable}{unstable}

\includegraphics[scale=0.32]{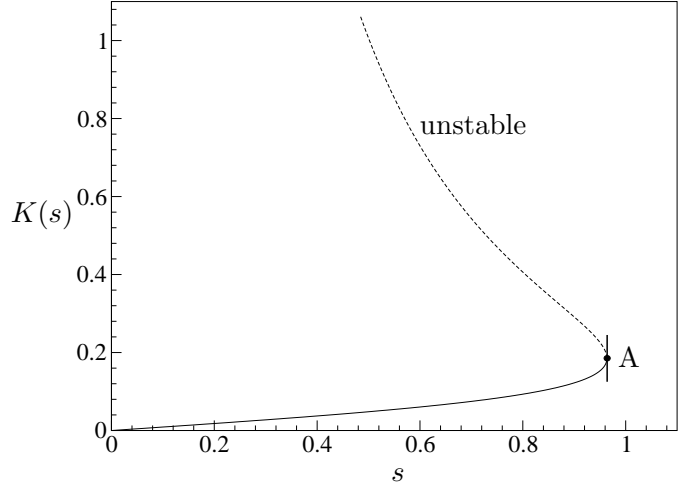}
\caption{The pair of conjugate variables $s$ and $K(s)$ for the RCR
disk. At the point A the parameter $s$ takes the \lq\lq critical\rq\rq\
value $s=0.9634\dots$ (corresponding to $b=1.3393\dots$).
The vertical tangent at that point indicates a change of stability.} 
\label{KonjPar}
\end{figure}

The numerical discussion of $\tilde E(b;s)$ leads to the
following results:
The extremum condition $\partial\tilde E/\partial b=0$ yields the
parameter relation \eqref{16b1}. That means the RCR disk solution is an
extremum of $\tilde E$ (at least a stationary point).
As it was shown by Poincar\'e and discussed by Katz in \cite{Katz},
stability can be analysed with the help of conjugate variables with
respect to a thermodynamic potential. Here, 
we may choose the variable $s$
and its conjugate $K(s)$ with respect to $\tilde E$,
\begin{equation}\label{16ba}
K(s):=\frac{\partial\tilde E}{\partial s}[b_\textrm{e}(s),s],
\end{equation}
where $b=b_\textrm{e}(s)$ is the equilibrium relation between $b$ and $s$.
Fig.~\ref{KonjPar} shows the pair $[s,K(s)]$. The criterion for a change
of stability is the existence of a vertical tangent to this
curve. Obviously, such a behaviour is given at the point A where
 $s=0.9634\dots$ ($b=1.3393\dots$). A careful numerical analysis of the
 curve $K(s)$ shows that there is no other point with a vertical
 tangent. Since a stable branch can be identified
 as the one with a
positive slope near the vertical tangent, we conclude that the
lower branch in Fig.~\ref{KonjPar} is stable according to Poincar\'e's
definition
and the upper branch (dashed curve) is
unstable:
\emph{RCR disks are unstable for $b>1.3393\dots$}  
This result includes the \emph{instabilty of RCR disks with negative
binding energy $E_\textrm{b}$} ($E_\textrm{b}<0$ for $b>4.1074\dots$).

Because of the definition \eqref{E9b},
$M_0^2/J$
takes its maximum just at the critical value
$b=1.3393\dots$ (see Fig.~\ref{figure2}).
\section{Particular collision processes}\label{conservation}

\subsection{Conservation laws}\label{conservationA}

As mentioned before, we consider \lq\lq head on\rq\rq\ collisions of
rigidly rotating disks of dust for parallel and antiparallel angular
momenta as sketched in Fig.~\ref{figure1} [from now on denoted as
scenario (a) and scenario (b), respectively].   
To compare the initial disks with the resulting merged disk we will make
use of conservation laws. Obviously, 
there are two conserved quantities. One of them
is the \emph{baryonic mass}. Considering two colliding bodies A and B,
we have
\begin{equation}\label{17x}
\tilde M_0=M_0^A+M_0^B,
\end{equation}
where the baryonic mass $\tilde M_0$ of the final body is
the sum of the 
baryonic masses $M_0^A$ and $M_0^B$ of the colliding bodies
(from now on, tildes denote
quantities of the final bodies).
A first consequence of \eqref{17x} is that the (inelastic) collision of
two bodies with positive binding energies cannot lead to a body with a
negative binding energy. Namely,
\begin{equation}\label{17a}
\tilde M_0-\tilde M = (M_0^A-M^A)+(M_0^B-M^B)+(M^A+M^B-\tilde M).
\end{equation}
According to our assumption, the first two terms on the right hand side
are positive. Due to the loss of energy by gravitational radiation the
last term has to be positive, too,
\begin{equation}\label{17aa}
\tilde M<M^A+M^B.
\end{equation}
Hence ($\tilde M_0>0$) the relative
binding energy of the resulting body is positive,
\begin{equation}\label{17b}
\tilde E_\textrm{b}=1-\frac{\tilde M}{\tilde M_0}>0.
\end{equation}
Applying this result to our disks of dust we may
exclude the \lq\lq strange\rq\rq\ branch (Fig.~\ref{Bindungsenergie},
$b>4.1074\dots$) 
of the RCR disk solution: RCR
disks with  $b>4.1074\dots$, $E_\textrm{b}<0$ cannot form by collisions.

From now on,
we confine ourselves to colliding RR disks with \emph{equal}
baryonic masses, $M_0^A=M_0^B=M_0$. Then, Eq.~\eqref{17x} takes the
form
\begin{equation}\label{17}
\tilde M_0=2M_0.
\end{equation}
(For dust, the conservation of the baryonic mass is a consequence of the
local energy-momentum conservation ${T^{ij}}_{;j}=0$.)

Due to the existence of an
azimuthal killing vector $\eta^i$ the \emph{angular momentum} is
conserved as well. In the first scenario, the two parallel angular
momenta $J$ of the two
rigidly rotating disks sum up to $2J$ in the final disk, 
\begin{equation}\label{18}
\tilde J=2J.
\end{equation}
In a second case we will study questions connected with the formation of
the RCR disk by collisions of two one-component disks with antiparallel
angular momenta as described in
section \ref{disk} and assume the
conservation of the angular momentum for
each component separately,
\begin{equation}\label{18a}
\pm\tilde  J=\pm J
\end{equation}
(the resulting angular momentum of the initial RR disks and the final RCR
disk vanishes). It should be emphasized that the assumption \eqref{18a}
is not very realistic. A small amount of friction between the two dust
components would violate the separate conservation of the angular
momenta.
Nevertheless an assumption like \eqref{18a} can lead to deeper insight
into the physical processes connected with the formation of
counter-rotating disks and allows, by way of example, a comparison
between collisions of disks with parallel and antiparallel angular
momenta.   

The conservation equations \eqref{17}-\eqref{18a} enable us to calculate
the parameters characterizing the
final disk as functions of the parameters of the initial RR
disks without studying
the intermediate, complicated
dynamical process.

\subsection{Collision of disks with parallel angular momenta}
The equations in section \ref{disk} show that the
combination $M_0^2/J$ of the conserved quantities depends on $\mu$
alone (and not on some second parameter). Therefore the relation
\begin{equation}\label{19a}
\frac{\tilde M_0^2}{\tilde J} \equiv \frac{M_0^2}{J}(\tilde\mu)=2\frac{M_0^2}{J}(\mu)
\end{equation}
combining the equations \eqref{17} and \eqref{18} allows us to calculate the
parameter $\tilde\mu$ of the final disk as a function of $\mu$
alone.
\begin{figure}
\psfrag{mu}{$\mu$}
\psfrag{MdJ1}{$\displaystyle\frac{M_0^2}{J}$}
\includegraphics[scale=0.32]{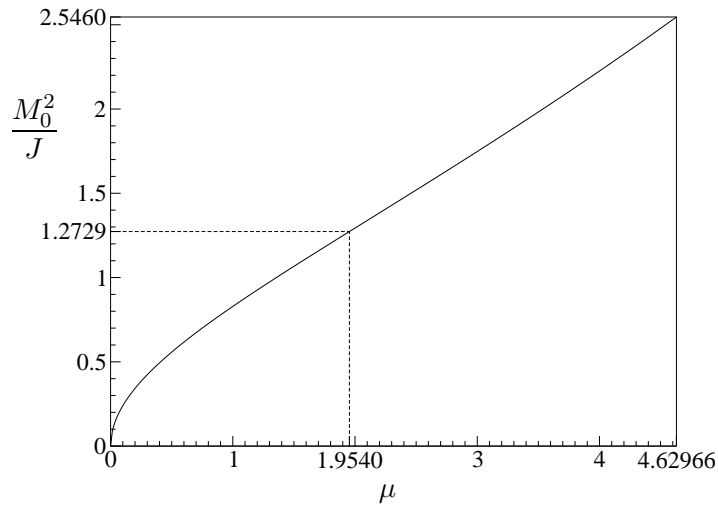}
\caption{$M_0^2/J$ of the rigidly rotating disk as function of the 
centrifugal parameter $\mu$}
\label{figure1a}
\end{figure}
Using the formulae \eqref{3} and \eqref{4} of appendix \ref{AppendixA}
Fig.~\ref{figure1a} shows that the initial disk ratio
$M_0^2/J$ as a function of $\mu$
reaches a maximum of
$\max\left(M_0^2/J\right)=2.5460\dots$ at the ultrarelativistic limit of the disk of dust
$\mu=4.62966\dots$ Thus Eq. \eqref{19a} cannot have solutions for all values
of $\mu$. Obviously, the solutions are restricted to the interval
\begin{equation}\label{19c}
0\le \frac{M_0^2}{J}\le \frac{1}{2}\max\left(\frac{M_0^2}{J}\right)=1.2729\dots
\end{equation}
that is equivalent to the interval
\begin{equation}\label{19b}
0\le\mu\le 1.9540\dots,
\end{equation}
cf. the firt picture of Fig.~\ref{figure3}.

\emph{Only for initial disks in this parameter range can the collision 
again lead to a rigidly rotating disk of dust}.
Beyond the limit
$\mu=1.9540\dots$ a collision must lead to other final states,
e.g. black holes or black holes surrounded by  matter rings.

On the other hand,
$\tilde\mu$ can take on all
values in the interval $[0,4.62966\dots]$ being considered, i.e. there is no
restriction on the parameters of the resulting RR disks. That means
\emph{every rigidly rotating disk can be formed in a collision of two rigidly
rotating initial disks}. If the centrifugal
parameter $\mu$ of the initial disks approaches the maximum $\mu=1.9540\dots$, then
the collision leads to a rigidly rotating disk with
$\tilde\mu=4.62966\dots$ arbitrarily close to the extreme Kerr black hole.  

Having solved  \eqref{19a} to obtain $\tilde\mu=\tilde\mu(\mu)$, cf. the
first graph of figure \ref{figure3},
one may calculate the new coordinate radius
$\tilde\rho_0$ from
\eqref{17} via $\tilde M(\tilde\mu,\tilde\rho_0)=2M_0(\mu,\rho_0)$.
The result is plotted in the second graph of figure \ref{figure3}.
\subsection{Collision of disks with antiparallel angular momenta}
\label{antiparallel}
As with the rigidly rotating disk, the ratio
$M_0^2/J$ for the counter-rotating disk depends only on a
centrifugal parameter, here $b$,
and not on $\rho_0$. With the definition
(cf. appendix \ref{AppendixA1})
\begin{equation}\label{19}
f(\tilde b):=\frac{\tilde M_0^2}{\tilde J}=
\frac{
\left[\ln\frac{\sqrt{1+4\tilde b^2}}{2}\arctan(2\tilde b)
-\Im\left(\mathrm{dilog}\frac{1+2\ii
\tilde b}{2}\right) \right]^2
}{
\pi\left[4\tilde b-\left(2+\ln\frac{\sqrt{1+4\tilde
b^2}}{2}\right)\arctan(2\tilde b)
+\Im\left(\mathrm{dilog}\frac{1+2\ii \tilde b}{2}\right) \right]
}
\end{equation}
and with the conservation equations \eqref{17} and \eqref{18a} one finds
\begin{equation}\label{20}
\frac{\tilde M_0^2}{\tilde J}(
\tilde b)\equiv f(\tilde b)=4\frac{M_0^2}{J}(\mu),
\end{equation}
which can be used to calculate the counter-rotating disk parameter $\tilde b$
as a function of the initial disk parameter $\mu$.

\begin{figure}
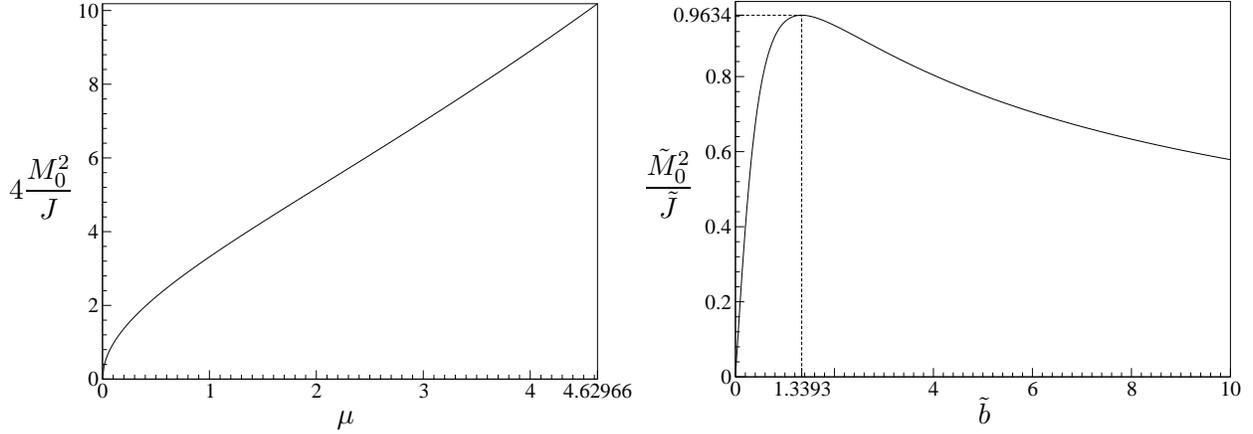

\psfrag{b}{$\tilde b$}
\psfrag{bmax}{$b_\textrm{max}$}
\psfrag{fmax}{$f_\textrm{max}$}
\psfrag{h(b)}{$\displaystyle\frac{\tilde M_0^2}{\tilde{J}}$}
\psfrag{mu}{$\mu$}
\psfrag{dJ}{$\displaystyle\!\! 4\frac{M_0^2}{J}$}
\includegraphics[scale=0.28]{fig5a.eps}
\includegraphics[scale=0.28]{fig5b.eps}
\caption{Baryonic mass $M_0$ and angular momentum $J$ before and after
the collision, cf. Eq.\eqref{20}.}
\label{figure2}
\end{figure}

Similar to the case of scenario (a), by using the formulae of the
appendices \ref{AppendixA} and \ref{AppendixA1}
it turns out, that \eqref{20} does not have solutions $\tilde b$ for all
values $\mu$ in the allowed range $0\le\mu\le 4.62966\dots$
Fig.~\ref{figure2}  shows that $f(\tilde b)$ reaches a maximum of
$f_\textrm{max}=0.96344\dots$ for $\tilde b=b_\textrm{max}=1.33934\dots$, while the right hand side of \eqref{20}
grows to much larger values. Thus the formation of a counter-rotating disk
is only possible in the small
range
\begin{equation}\label{20a}
0\le \frac{M_0^2}{J}\le \frac{1}{4}f_\textrm{max}=0.2408\dots,
\end{equation}
i.e. from initial disks in the parameter range
\begin{equation}\label{21}
0\le\mu\le 0.101003\dots
\end{equation}
For values of $\mu$ in this interval Eq. \eqref{20} always has two
solutions: $\tilde b$ can be smaller or larger than $b_\textrm{max}$,
$\tilde b \lessgtr
b_\textrm{max}$.
Later it will turn out, that in general
only the solution $\tilde b<b_\textrm{max}$ is of physical relevance,
since the formation of a counter-rotating disk system with the larger
value of $\tilde b$ would only be possible if
external energy were put into the
system.

Finally, if $\tilde b$ has been calculated, \eqref{17} can be used to
determine $\tilde\rho_0$.

\section{Properties of the merged disks}\label{properties}
\subsection{General parameter conditions}\label{numeric}
\begin{figure}
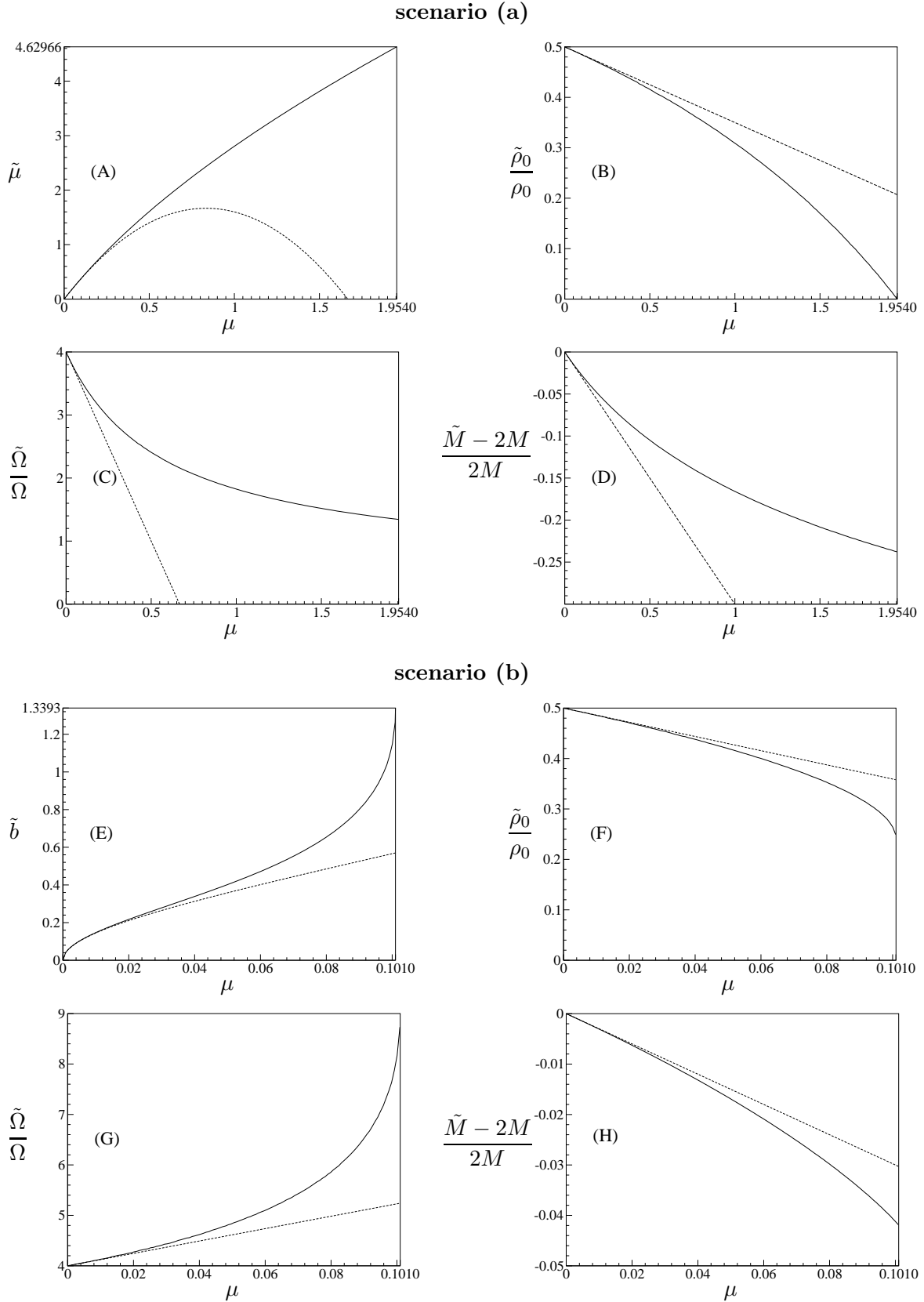

\psfrag{mu}{$\mu$}
\psfrag{mu1}{$\tilde\mu$}
\psfrag{b}{$\tilde b$}
\psfrag{Ome}{$\displaystyle\frac{\tilde\Omega}{\Omega}$}
\psfrag{rho}{$\displaystyle\frac{\tilde\rho_0}{\rho_0}$}
\psfrag{E}{$\hspace{-1.2cm}\displaystyle\frac{\tilde M-2M}{2M}$}
{\bf scenario (a)}\\[1.5ex]
\includegraphics[scale=0.24]{fig6a.eps}\qquad\qquad
\includegraphics[scale=0.24]{fig6b.eps}\\[1.5ex]
\includegraphics[scale=0.24]{fig6c.eps}\qquad\qquad
\includegraphics[scale=0.24]{fig6d.eps}\\
{\bf scenario (b)}\\[1.5ex]
\includegraphics[scale=0.24]{fig6e.eps}\qquad\qquad
\includegraphics[scale=0.24]{fig6f.eps}\\[1.5ex]
\includegraphics[scale=0.24]{fig6g.eps}\qquad\qquad
\includegraphics[scale=0.24]{fig6h.eps}
\caption{Parameters of the
resulting disks after collision (tilded quantities)
as functions of the initial disk
parameter $\mu$ in the allowed range
$0\le\mu\le 1.9540\dots$ [scenario (a)] and
$0\le\mu\le 0.101003\dots$ [scenario (b)]
compared with the post-Newtonian approximations
for small $\mu$ (dashed lines), cf. section \ref{approximation}.
$\mu$, $\Omega$, $\rho_0$ and $M$ belong to the initial disks.
}
\label{figure3}
\end{figure}

In the last section we showed which initial parameters 
can lead to rigidly rotating or counter-rotating disks after a collision. 
Now we will discuss the parameters of the resulting disks as functions
of the initial parameters. For this purpose we make use of the explicit
expressions for mass $M$, baryonic mass $M_0$ and angular momentum $J$, 
\eqref{2}-\eqref{4} and \eqref{14}-\eqref{16}, and plug them into the
conservation laws \eqref{19a}, \eqref{20} and \eqref{17}. In this way we
obtain the desired parameter relations implicitly in terms of elliptic
functions. In Fig.~\ref{figure3} one can see a number of parameter
relations generated by a numerical evaluation of those implicit
relations (solid lines). They may be compared with
the
corresponding
post-Newtonian approximations derived in subsection \ref{approximation}
(dotted lines). As
expected, the curves
agree well for small values of
the centrifugal parameter $\mu$, but differ for larger $\mu$ even qualitatively.  

As mentioned earlier [Eq. \eqref{19b}], in scenario (a) the formation of an RR disk after
the collision is only possible for $0\le\mu\le 1.9540\dots$ while for
the final disk all parameters $0\le\tilde\mu\le 4.62966\dots$ are
allowed (Fig.~\ref{figure3}A).
The new coordinate radius $\tilde\rho_0$ is one half of $\rho_0$ in the
Newtonian limit, but smaller in the general case. In the limit
$\mu=1.9540\dots$ the collision leads to a disk in transition to
the extreme Kerr black hole
with vanishing coordinate radius, $\tilde\rho_0=0$ (Fig.~\ref{figure3}B).
The angular velocity increases, maximally by a
factor of four in the Newtonian limit (Fig.~\ref{figure3}C).
\emph{For the maximum amount of lost energy $\tilde M-2M$
during the merger process due to gravitational waves one obtains a
limit of $23.8\%$ of the total initial mass} (Fig.~\ref{figure3}D).

\begin{figure}
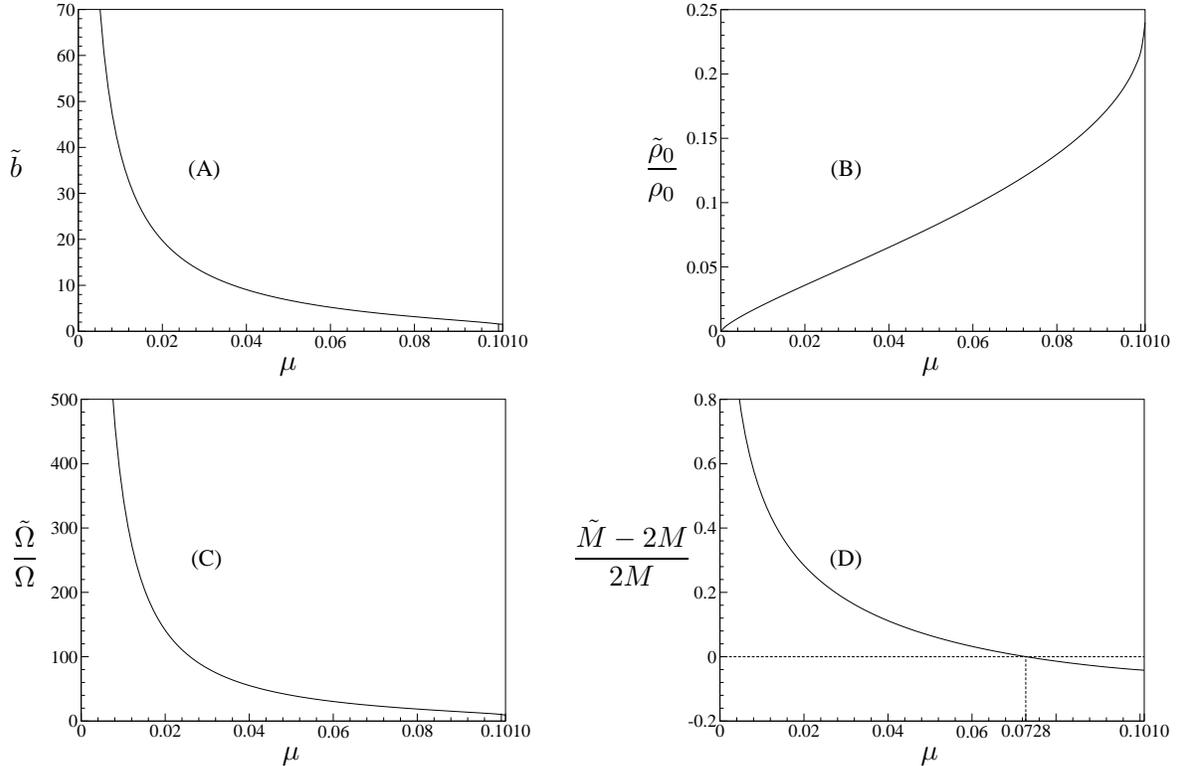

\psfrag{mu}{$\mu$}
\psfrag{b}{$\tilde b$}
\psfrag{O}{$\displaystyle\frac{\tilde\Omega}{\Omega}$}
\psfrag{rho}{$\displaystyle\frac{\tilde\rho_0}{\rho_0}$}
\psfrag{M}{\hspace{-1cm}$\displaystyle\frac{\tilde M-2M}{2M}$}
\includegraphics[scale=0.24]{fig7a.eps}\qquad\qquad
\includegraphics[scale=0.24]{fig7b.eps}\\[1.5ex]
\includegraphics[scale=0.24]{fig7c.eps}\qquad\qquad
\includegraphics[scale=0.24]{fig7d.eps}
\caption{Behaviour of the parameters of the resulting RCR disks
[scenario (b)] 
for the solutions
$\tilde b>b_\textrm{max}$.}
\label{figure4}
\end{figure}

In scenario (b) the formation of an RCR disk is only possible in the small
parameter range $0\le\mu\le 0.101003\dots$  As mentioned before
(Sec. \ref{antiparallel}), there
always exist two values of $\tilde b$,
$\tilde b
\lessgtr
b_\textrm{max}=1.33934\dots$,
for a given
$\mu$ (or $M_0^2/J$).
In the first case ($\tilde b<b_\textrm{max}$), the coordinate radius
$\tilde\rho_0$ shrinks to
one half of the initial radius or smaller,
but never reaches zero (Fig.~\ref{figure3}F). The angular
velocity increases by a factor of four or larger (Fig.~\ref{figure3}G)
 and \emph{the limit for the
relative energy loss is $4.2\%$}  (Fig.~\ref{figure3}H). In the second
case ($\tilde b>b_\textrm{max}$),
the new coordinate
radius $\tilde\rho_0$ is arbitrarily small for small $\mu$
(Fig.~\ref{figure4}B) and the
angular velocity grows to infinity for $\mu\to 0$ (Fig.~\ref{figure4}C).

We have seen that the emission condition \eqref{17aa}
\begin{equation}\label{Zus1}
\tilde M<M^A+M^B=2M
\end{equation}
forbids the formation of RCR disks with negative binding energy
($\tilde b>4.1074\dots$). The preceding analysis together with the exclusion of
energy transfer into the system \eqref{Zus1} \emph{enlarges} the forbidden
$\tilde b$-interval and \emph{restricts} the formation of RCR disks
aditionally. 
Figs. \ref{figure3}H and \ref{figure4}D show that Eq. \eqref{Zus1}
is completely
satisfied for the first interval
($0<\tilde b\le b_\textrm{max}=1.33934\dots$) and
holds for the small piece $0.0728\dots\mbox{}\!\le\mu\le 0.101003\dots $
of the second intervall corresponding to
$1.33934\dots\mbox{}\!\le\tilde b\le 3.8038\dots $ but is violated for
$0\le\mu< 0.0728\dots$ which corresponds to
$3.8038\dots\mbox{}<\tilde b<\infty$.
Obviously, this implies our former result that RCR disks formed by
collisions have always a positive binding energy, $\tilde
E_\textrm{b}>0$ for
$\tilde b<4.1074\dots$

Adding now the result of
section~\ref{EquilibriumStability} where we showed the instability of
RCR disks in the interval $b>1.3393\dots$ we
conclude that for physical reasons
\emph{we cannot expect the formation of RCR
disks in the second branch $b>1.3393\dots$}
\subsection{Analytic parameter conditions for weakly relativistic disks}\label{approximation}
For most astrophysical applications, rigidly rotating disk models
are characterized by very small values of  the
centrifugal
parameter, $\mu\ll 1$. For such weakly relativistic disks it is useful
and possible to obtain
analytic parameter conditions derived by
post-Newtonian expansions of the equations of state and the
conservation laws \eqref{19a} and \eqref{20}. 
(For counter-rotating disks, for which
Eq. \eqref{20} has two
solution branches,
we restrict ourselves to solutions with $\tilde b\ll
b_\textrm{max}$ excluding the \lq\lq strange\rq\rq\
branch with negative binding energy.)
\subsubsection{Scenario (a)}
The expansion of the relations in appendix \ref{AppendixA} in a power
series for $\mu$ leads to
\begin{equation}\label{22}
\Omega
M=\frac{\sqrt{2}}{\pi}\left(\frac{1}{3}\mu^{3/2}-\frac{1}{10}\mu^{5/2}\right)+\mathcal
O(\mu^{7/2}),
\end{equation}
\begin{equation}\label{23}
\Omega
M_0=\frac{\sqrt{2}}{3\pi}\left(\mu^{3/2}-\frac{1}{5}\mu^{5/2}\right)+\mathcal
O(\mu^{7/2}),
\end{equation}
and
\begin{equation}\label{24}
\Omega^2
J=\frac{\sqrt{2}}{15\pi}\left(\mu^{5/2}-\frac{1}{2}\mu^{7/2}\right)+\mathcal
O(\mu^{9/2}).
\end{equation}
The conservation equation \eqref{19a} now can be written as
\begin{equation}\label{24a}
10\tilde\mu^{1/2}+\tilde\mu^{3/2}+\mathcal
O(\tilde\mu^{5/2})=2(10\mu^{1/2}+\mu^{3/2})+\mathcal O(\mu^{5/2})
\end{equation}
with the solution
\begin{equation}\label{24b}
\tilde\mu=4\mu-\frac{12}{5}\mu^2+\mathcal O(\mu^3).
\end{equation}
With $\tilde\mu$ and the expansions \eqref{23} and
$\omega=\Omega\rho_0=1/\sqrt{2}\,(\mu^{1/2}-1/2\,\mu^{3/2})+\mathcal O(\mu^{5/2})$
one can calculate $\tilde\rho_0$ from Eq. \eqref{17},
\begin{equation}\label{24c}
\tilde\rho_0/\rho_0=\frac{1}{2}-\frac{3}{20}\mu+\mathcal O(\mu^2).
\end{equation}
That means, in the lowest order, the new coordinate radius is one half of the
original radius.
For the change of the angular velocity, we obtain
\begin{equation}\label{24d}
\tilde\Omega/\Omega=4-6\mu+\mathcal O(\mu^2),
\end{equation}
i.e. the rotation of the final disk
will be four times faster than the rotation of the initial
disks
in the lowest order.
By comparing the gravitational masses of the two initial disks and the
final disk, one can determine the energy loss
due to gravitational radiation during the dynamical process. The
relative energy loss is
\begin{equation}\label{24e}
(\tilde M-2M)/2M=-\frac{3}{10}\mu+\mathcal O(\mu^2),
\end{equation}
where the minus sign shows, that the energy indeed leaves the system. 

To compare the post-Newtonian approximations with
the exact parameter conditions of the last section
they are plotted in figure \ref{figure3} as dotted curves.

\subsubsection{Scenario (b)}
With the expansion of Eq. \eqref{19} in terms of the centrifugal
parameter $\tilde b$, $f(\tilde
b)=\frac{4}{3\pi}(5\tilde b-\frac{64}{7}\tilde b^3)+\mathcal O(\tilde b^5)$,
Eq. \eqref{20} takes the form
\begin{equation}\label{25}
(10\mu^{1/2}+\mu^{3/2})+\mathcal O(\mu^{5/2})=\sqrt{2}(5\tilde
b-\frac{64}{7}\tilde b^3)+\mathcal O(\tilde b^5)
\end{equation}
with the solution
\begin{equation}\label{26}
\tilde b=\sqrt{2}\left(\mu^{1/2}+\frac{263}{70}\mu^{3/2}\right)+\mathcal O(\mu^{5/2}).
\end{equation}
From this and with $\tilde
M_0=4\tilde\rho_0/3\pi\left(\tilde b^2-\frac{11}{5}\tilde
b^4+\mathcal O(\tilde b^6)\right)$ one finds for the
coordinate radius of the disk
\begin{equation}\label{27}
\tilde\rho_0/\rho_0=\frac{1}{2}-\frac{197}{140}\mu+\mathcal O(\mu^2).
\end{equation}
The change of the angular velocity can be calculated from
$\Omega=\omega/\rho_0$
 and $\tilde\Omega=\tilde b/(\tilde\rho_0\sqrt{1+4\tilde
b^2})$. It leads to
\begin{equation}\label{28}
\tilde\Omega/\Omega=4+\frac{86}{7}\mu+\mathcal O(\mu^2).
\end{equation}
For the relative energy loss one obtains
\begin{equation}\label{29}
(\tilde M-2M)/2M=-\frac{3}{10}\mu+\mathcal O(\mu^2).
\end{equation}

Comparing the collisions of disks with parallel and antiparallel angular
momenta,
it turns out, that to the lowest order, the change of the angular
velocity and the change of the
coordinate radius are equal in both scenarios. Namely,
the lowest order terms represent the Newtonian results
and in the Newtonian theory there is no influence of the direction of
rotation on the gravitational field.
Interestingly, a similar effect occurs in the post-Newtonian regime: 
The energy loss for both scenarios also coincides to the
lowest order.

As an example, one can calculate the energy loss for the merger of two
disks with mass and radius of our milky way ($\rho_0=15000\,\mathrm{pc}$ and
$M=2\cdot 10^{11}M_\odot$, respectively). The corresponding centrifugal parameter
$\mu=3\cdot 10^{-6}$  is so small that the post-Newtonian approximation
is sufficient to calculate the efficiency.
With Eq. \eqref{29} one finds for the energy loss
$\Delta M=\tilde M-2M=-4\cdot 10^5 M_\odot$ for both scenarios.


\section{The efficiency of gravitational emission}\label{efficiency}
An important quantity in the collision process is its efficiency
$\eta$, which is the negative value of the relative energy loss,
\begin{equation}\label{34}
\eta=\frac{2M-\tilde M}{2M}.
\end{equation}
Obviously, $\eta$ is the \emph{upper limit} for the energy
transported away by gravitational radiation.
(Some part of the energy will also dissipate due
to friction. To guarantee the formation of a new rigidly
rotating or rigidly counter-rotating disk, friction is necessary to
force a constant angular velocity.)

As shown in section \ref{approximation}, for weakly relativistic disks
the efficiency is given as
\begin{equation}\label{34a}
\eta\approx 0.3\mu,
\end{equation}
cf. \eqref{24e}.

In section
\ref{numeric} we already presented the relative energy loss
as a function of the
centrifugal parameter $\mu$ of the initial disks,
cf. Fig.~\ref{figure3}D.
Here we derive a more explicit expression for the efficiency
making use of the scaling behaviour of the disks: $M/M_0$ and
$M_0^2/J$ are functions of the centrifugal parameter
alone ($\mu$ for RR and $b$ for RCR).
Therefore $M/M_0$ is a function of $M_0^2/J$,
\begin{equation}\label{34b}
\frac{M}{M_0}=F(x),\quad x:=\frac{M_0^2}{J}.
\end{equation} 
Using the conservation laws \eqref{17}
and \eqref{18} or \eqref{18a}  we obtain the following expressions
for the efficiency $\eta$ as a function of the quantity
$M_0^2/J$ of the initial disks,
\begin{equation}\label{35}
\eta(x)=1-\frac{F(2x)}{F(x)} \qquad \textrm{for scenario (a)},
\end{equation}
\begin{equation}\label{36}
\eta(x)=1-\frac{\tilde F(4x)}{F(x)} \qquad \textrm{for scenario (b)},
\end{equation}
where $F$ and $\tilde F$ can be calculated in terms of elliptic
functions from the equations of state
of the RR disks and the RCR disks, respectively. 
Accordingly, $\eta$ is determined completely by the relative binding
energy $E_\textrm{b}=1-M/M_0=1-F(x)$ of the RR or RCR disks.
\begin{figure}
\psfrag{eta}{$\eta$}
\psfrag{M0qdJ}{$M_0^2/J$}
\includegraphics[scale=0.32]{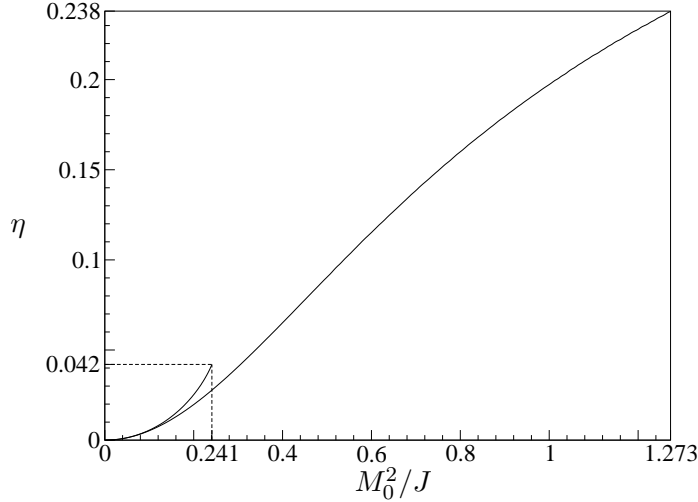}
\caption{The efficiency $\eta$ for scenario (a) (long curve) and
scenario (b) (short curve) as function of the physical quantity
$M_0^2/J$ of the initial RR disks up to the natural
end points
of the intervals for which disk formation is possible.}
\label{figure5}
\end{figure}

The efficiencies \eqref{35} and \eqref{36} are plotted in
Fig.~\ref{figure5}. Note that the efficiency measures the energy
\emph{loss}. The picture shows that
the formation of a counter-rotating RCR disk [scenario (b)] is
\lq\lq more efficient\rq\rq\ than the formation of a rigidly rotating disk [scenario
(a)]
for initial disks with the same amount of angular momentum and baryonic
mass,
but is
possible only in a much smaller parameter range. It ends with a
comparably small maximum value of $\eta\approx 4.2\%$.
The limit $\eta\approx 23.8\%$ for the collision of disks with parallel
angular momenta has the order of magnitude of
Hawking's and Ellis' limit $\eta\approx 29.3\%$ for the coalescence of two
spherically symmetric black holes.
\section{Conclusion}

In this paper we have
discussed the collision and merger processes of rigidly rotating
(RR) disks leading again to a RR disk or a rigidly counter-rotating (RCR) disk.
The conservation equations for mass and angular momentum
showed, that these processes are restricted to limited parameter
intervals for the initial disks: Only for
rigidly rotating disks whose centrifugal parameter $\mu$
does not exceed a given maximum value do these conditions allow the
formation of a new rigidly rotating disk, and in an even much smaller
parameter range, the process can lead to a counter-rotating disk. (But of
course, even if the balance equations can be satisfied in these
parameter ranges, it is not clear if these processes are 
dynamically possible.)
We were able to calculate
the physically relevant relations between the parameters (\lq\lq
equations of state\rq\rq)
by a numerical evaluation of the exact but
implicit conditions resulting from the conservation equations.
Explicit analytical expressions could be derived for the post-Newtonian domain.

It turned out, that every RR disk can be the result of such a collision
process (every point of the two-dimensional parameter range can be
reached), while the formation of RCR disks is restricted. RCR disks with
negative binding energy (as described by a branch of the RCR solution)
cannot be formed in a collision process.
An interesting result is the calculated upper limit for the
efficiency of $\eta\approx 23.8\%$ for the formation of RR disks and of
 $\eta\approx 4.2\%$ for the formation of RCR disks.

Counter-rotating disks are interesting initial
configurations for axisymmetric collapse scenarios with gravitational
emission. 
In these cases, the mathematical analysis consists in the discussion of
initial-boundary problems for the vacuum Einstein equations. A typical
example will be published elsewhere.


\begin{acknowledgments}
We would like to thank David Petroff for many valuable discussions and
the referee for his suggestion to perform a stability analysis.
This work was funded by SFB/Transregio 7 \lq\lq Gravitationswellenastronomie\rq\rq.
\end{acknowledgments}

\appendix

\section{Why rigid rotation?}\label{Extremal}
If there is a small amount of friction between the dust rings (which can
be anticipated for every realistic system),
differentially rotating
disks will develop towards an equilibrium state with rigid rotation. Due
to the loss of energy via gravitational or thermal radiation the
gravitational mass $M$ of the disk decreases in these processes.
Therefore we expect to
find rigidly rotating disks as minima of the gravitational mass $M$
compared to differentially rotating disks with the same baryonic mass
$M_0$ and the same angular momentum $J$. Here, we show that $M$ indeed
takes an extremum.
\footnote{To avoid confusion we want to point out that we
use two different kinds of variational principles in this paper. In this
section, we
compare different \emph{solutions} to the Einstein equations
(with different rotational laws) to find the
rigidly rotating disks as extrema of the gravitational mass $M$.
In contrast to this we
consider \emph{trial functions} that do \emph{not} solve the field
equations in Sec.~\ref{EquilibriumStability} to find the solutions as
extrema of the thermodynamic potential $E$.}

Following the calculations in \cite{NeugebauerTD}, one can generalize
the Gibbs formula \eqref{1c0} to disks with differential rotation
$\Omega=\Omega(\rho)$. The result is
\begin{equation}\label{e1}
\delta
M=\int\limits_{t=t_0}[\mu_0\delta(\rho_{M_0}\sqrt{-g})+\Omega\delta(\rho_J\sqrt{-g})]\,\dd^3x,
\end{equation}
where $\mu_0=\mu_0(\rho)$ is the chemical potential (specific free
enthalpy),
$g$ is the
determinant of the metric tensor  and $\rho_{M_0}$ and
$\rho_J$ denote the densities of baryonic mass and angular momentum,
respectively, defined by
\begin{equation}\label{e2}
M_0=\int\limits_{t=t_0}\rho_{M_0}\sqrt{-g}\,\dd^3x,\quad
J=\int\limits_{t=t_0}\rho_J\sqrt{-g}\,\dd^3x.
\end{equation}
If one now considers the set of all disks with different rotation laws
$\Omega=\Omega(\rho)$, but fixed baryonic mass and angular momentum,
i.e. with the constraints
\begin{equation}\label{e3}
C_1:=M_0=\textrm{constant},\quad C_2:=J=\textrm{constant},
\end{equation}
then one can find the disks with extremal gravitational mass $M$
(and
accordingly 
--- due to $M_0=\textrm{constant}$ ---  with extremal binding energy
$M_0-M$). With the Lagrange multipliers $\lambda_1$ and $\lambda_2$, the
condition for a stationary point is
\begin{equation}\label{e4}
\delta(M+\lambda_1 C_1+\lambda_2 C_2)=0.
\end{equation}
Using \eqref{e1}-\eqref{e3}, the latter equation takes the form
\begin{equation}\label{e5}
\int\limits_{t=t_0}[(\mu_0+\lambda_1)\delta(\rho_{M_0}\sqrt{-g})
+(\Omega+\lambda_2)\delta(\rho_J\sqrt{-g})]\,\dd^3x=0.
\end{equation}
Since $\rho_{M_0}$ and $\rho_J$ are varied independently, one finds
\begin{equation}\label{e6}
\Omega=\textrm{constant},\quad \mu_0=\textrm{constant},
\end{equation}
i.e. the family of rigidly rotating disks of dust.

The same considerations are valid for counter-rotating disks of
dust. The generalization of Eq. \eqref{e1} is
\begin{equation}\label{e7}
\delta
M=\sum\limits_{i=1}^2\int\limits_{t=t_0}[\stackrel{(i)}{\mu_0}\delta(\stackrel{(i)}{\rho_{M_0}}\sqrt{-g})+\stackrel{(i)}{\Omega}\delta(\stackrel{(i)}{\rho_J}\sqrt{-g})]\,\dd^3x,
\end{equation}
where the index $i$ distinguishes between the two fluid components. With the
restriction to disks with the same chemical potential
($\stackrel{(1)}{\mu_0}=\stackrel{(2)}{\mu_0}=:\frac{1}{2}\mu_0$)
and baryonic
mass density
($\stackrel{(1)}{\rho_{M_0}}=\stackrel{(2)}{\rho_{M_0}}
=:\frac{1}{2}\rho_{M_0}$)
and  opposite values of angular velocity
($\stackrel{(1)}{\Omega}=-\stackrel{(2)}{\Omega}=:\Omega$) and
angular momentum density 
($\stackrel{(1)}{\rho_J}=-\stackrel{(2)}{\rho_J}=:\rho_J$),
Eq. \eqref{e7} reduces to Eq. \eqref{e1}, i.e. the extremal
problem leads to the \emph{rigidly} counter-rotating disks with $\Omega=\textrm{constant}$.

\section{Multipole moments of the rigidly rotating disk}\label{AppendixA}

By specifying the formulae of the general solution to the axis of
symmetry, it is possible to calculate all multipole moments of the disk
\cite{Kleinwaechter}. In particular, one can derive expressions for the
gravitational (ADM) mass $M$, the baryonic mass $M_0$ and the angular
momentum $J$,
\begin{equation}\label{2}
\Omega M(\mu)=-\frac{1}{2}\left(\omega(\mu)a_1(\mu)+b_0(\mu)\right),
\end{equation}
\begin{equation}\label{3}
\Omega M_0(\mu)=\frac{\sqrt{2\mu}}{4}a_1(\mu),
\end{equation}
\begin{equation}\label{4}
\Omega^2 J(\mu)=-\frac{1}{2}\left(\omega(\mu)
a_1(\mu)+\frac{1}{2}b_0(\mu)\right), 
\end{equation}
where one needs the relations
\begin{eqnarray*}
 a_1 & = &
 \frac{1}{\sqrt{\mu}}\Big[2\sqrt{1+\mu^2}I_0(\mu)\Big({h'}^2
 -\frac{E(h)}{K(h)}\Big)+I_1(\mu)
 +(1+\mu^2)^{1/4}\frac{\pi}{K(h)}\Lambda_0(\mathrm{am}
 (\hat I(\mu),h'),h))\Big]\\
 & \equiv & \frac{1}{\sqrt{\mu}}\Big[\sqrt{\frac{2}{h h'}}\left(E[\mathrm{am}(\hat
 I(\mu),h'),h']-h^2\hat I\right)+I_1(\mu)\Big],
\end{eqnarray*}
\begin{equation*}
 b_0=-\frac{1}{h(\mu)}\mathrm{sn}(\hat I(\mu),h'(\mu))\mathrm{dn}(\hat
   I(\mu),h'(\mu)),
\end{equation*}
\begin{equation*}
 \omega=\Omega\rho_0=\frac{1}{2}\sqrt{1-\frac{h'^2(\mu)}{h^2(\mu)}}\mathrm{cn}(\hat I(\mu),h'(\mu)) ,
\end{equation*}
\begin{equation*}
\ee^{2V_0}=\frac{h'(\mu)}{h(\mu)}\mathrm{cn}^2(\hat I(\mu),h'(\mu)),
\end{equation*}
\begin{equation*}
 h=\sqrt{\frac{1}{2}\left(1+\frac{\mu}{\sqrt{1+\mu^2}}\right)},\quad
 h'=\sqrt{1-h^2},
\end{equation*}
\begin{equation*}
 I_n=\frac{1}{\pi}\int\limits_0^\mu\frac{\ln(\sqrt{1+x^2}+x)}{\sqrt{1+x^2}}\frac{x^n}{\sqrt{\mu-x}}\dd x,\quad
 \hat I=(1+\mu^2)^{1/4}I_0(\mu),
\end{equation*}
with the complete elliptic integrals $K(k)=F(\pi/2,k)$ and $E(k)=E(\pi/2,k)$,
Heumann's Lambda function
$\Lambda_0(\psi,k)=\frac{2}{\pi}[E(k)F(\psi,k')+K(k)E(\psi,k')-K(k)F(\psi,k')]$,
$k'=\sqrt{1-k^2}$
and the Jacobian elliptic functions $\mathrm{sn}$, $\mathrm{cn}$, $\mathrm{dn}$ and  $\mathrm{am}$.

\section{Multipole moments of the rigidly counter-rotating disk}\label{AppendixA1}
For the metric \eqref{5},
by using the field equations, the relations \eqref{1a}-\eqref{1c} for mass and angular
momentum can be expressed using only
the metric potential $U$ and its $\rho$- and $\zeta$-derivative in
the disk. For general counter-rotating disks one finds
\begin{equation}\label{9}
M=\int\limits_0^{\rho_0}U_{,\zeta}\rho\,\dd\rho,
\end{equation}
\begin{equation}\label{10}
M_0=\int\limits_0^{\rho_0}\ee^{-U}\sqrt{(1-2\rho U_{,\rho})(1-\rho U_{,\rho})}U_{,\zeta}\rho\,\dd\rho
\end{equation}
and
\begin{equation}\label{11}
J=\frac{1}{2}\int\limits_0^{\rho_0}\ee^{-2U}\sqrt{\rho U_{,\rho}(1-\rho U_{,\rho})}U_{,\zeta}\rho^2\,\dd\rho.
\end{equation}
These equations can be
applied to rigidly counter-rotating disks (RCR disks) with the explicit
expressions (see \cite{Morgan})
\begin{equation}\label{12}
U(\rho,\zeta=0)=\frac{1}{2}\ln\left[
\frac{\Omega\rho_0}{2b}\left(1+\sqrt{1+4b^2\rho^2/\rho_0^2}\right)
\right],
\end{equation}
\begin{equation}\label{13}
U_{,\zeta}(\rho,\zeta=0+)=\frac{2b}{\pi\rho_0}\frac{1}{\sqrt{1+4b^2\rho^2/\rho_0^2}}\arctan\frac{2b\sqrt{1-\rho^2/\rho_0^2}}{\sqrt{1+4b^2\rho^2/\rho_0^2}},
\end{equation}
where $\Omega=b/(\rho_0\sqrt{1+4b^2})$ is the constant angular
velocity. They lead to
\begin{equation}\label{14}
M=\frac{\rho_0}{\pi}\left(1-\frac{1}{2b}\arctan(2b)\right),
\end{equation}
\begin{equation}\label{15}
M_0 
=  \frac{\rho_0}{4\pi b}
(1+4b^2)^{1/4}\left[\ln\frac{\sqrt{1+4b^2}}{2}\arctan(2b)
-\Im\left(\mathrm{dilog}\frac{1+2\ii b}{2}\right) \right]
\end{equation}
and
\begin{equation}\label{16}
J =  \frac{\rho_0^2}{16\pi
b^2}\sqrt{1+4b^2}\left[4b-\left(2+\ln\frac{\sqrt{1+4b^2}}{2}\right)\arctan(2b)
+\Im\left(\mathrm{dilog}\frac{1+2\ii b}{2}\right) \right]
\end{equation}
with the dilogarithm function
\begin{equation}\label{16aa}
\mathrm{dilog}(z)=\int\limits_1^z\frac{\ln w}{1-w}\,\dd w.
\end{equation}

\section{The counter-rotating disk of dust metric}\label{AppendixB}

The Inverse Scattering Method (cf. \cite{ISM}) can be used to calculate
the function $U(\rho,\zeta)$ in the line element \eqref{5} of the
counter-rotating RCR disk. For an angular velocity $\Omega$ and a coordinate
radius $\rho_0$ it leads to
\begin{equation}\label{A1}
U(\rho,\zeta)=\frac{1}{4\pi}\int\limits_{-1}^{1}\frac{\ln[1-\alpha(1-k^2)]}{\sqrt{(\ii
k-\zeta/\rho_0)^2+\rho^2/\rho_0^2}}\,\dd k,\quad \alpha:=4\Omega^2\rho_0^2.
\end{equation}
By using elliptical coordinates ($\xi,\eta$) with
\begin{equation}\label{A2}
\rho=\rho_0\sqrt{(1+\xi^2)(1-\eta^2)},\quad \zeta=\rho_0\xi\eta,
\end{equation}
it is possible to express $U$ in terms of the dilogarithm function \eqref{16aa}.
The result is
\begin{eqnarray}\label{A4}
U & = & \frac{1}{2\pi}\arctan\frac{1}{\xi}\cdot\ln\left[\frac{\alpha}{4}(1+\xi^2)(1+\eta)^2\right]-\frac{1}{2}\mathrm{Artanh}\,\eta-\frac{1}{2}\ln
n \nonumber\\
&& -\frac{1}{2\pi}\Im\sum_{j=1}^2[\mathrm{dilog}\,A^+_j+\mathrm{dilog}\,A^-_j],
\end{eqnarray}
where
\begin{equation}\label{A5}
A^{\pm}_j=1-\frac{{\beta}(1-\eta)(\xi+\ii)}{(-1)^j\sqrt{1+\beta^2(1+\xi^2-\eta^2)\pm
2\beta\xi\eta}-(\beta\xi\eta \pm 1)},
\quad \beta:=\sqrt{\frac{\alpha}{1-\alpha}}
\end{equation}
and
\begin{equation}\label{A6}
n=\frac{\sqrt{1+\beta^2(1+\xi^2-\eta^2)+2\beta\xi\eta}-1-\beta\xi\eta }{\beta\sqrt{(1+\xi^2)(1-\eta^2)}}.
\end{equation}
The function $k$ can be calculated from $U$ via line integration by
using the field equations
\begin{equation}\label{A7}
k_{,\rho}=\rho(U_{,\rho}^2-U_{,\zeta}^2), \quad k_{,\zeta}=2\rho U_{,\rho}U_{,\zeta}.
\end{equation} 


\begin{thebibliography}{99}

\bibitem{Neugebauer1} G. Neugebauer and R. Meinel, Astrophys. J. {\bf
414}, L97 (1993).

\bibitem{Neugebauer2} G. Neugebauer and R. Meinel, Phys. Rev. Lett. {\bf
73}, 2166 (1994). 

\bibitem{Neugebauer3} G. Neugebauer and R. Meinel, Phys. Rev. Lett.  {\bf
75}, 3046 (1995). 

\bibitem{Bardeen1} J. M. Bardeen and R. V. Wagoner, Astrophys. J. {\bf
158}, L65 (1969).

\bibitem{Bardeen2} J. M. Bardeen and R. V. Wagoner, Astrophys. J. {\bf
167}, 359 (1971).

\bibitem{Morgan} T. Morgan and L. Morgan, Phys. Rev. {\bf 183}, 1097
(1969); Errata: {\bf 188} 2544 (1969).

\bibitem{Bicak} J. Bi\v{c}\'{a}k, D. Lynden-Bell and J. Katz, Phys. Rev. D {\bf 47}, 4334 (1993).

\bibitem{Meinhardt} W. Meinhardt,
Ph.D. thesis,
Friedrich-Schiller-Universit\"at Jena, 1994.

\bibitem{Hawking} S. W. Hawking and G. F. R. Ellis, {\it The large scale
structure of space-time} ({\it Cambridge Monographs on Mathematical
Physics}) (Cambridge University Press, 1973).

\bibitem{Neugebauer4} G. Neugebauer, A. Kleinw\"achter and R. Meinel,
Helv. Phys. Acta {\bf 69}, 472 (1996).

\bibitem{Hartle} J. B. Hartle and D. H. Sharp, ApJ {\bf 147}, 317 (1967).

\bibitem{NeugebauerTD} G. Neugebauer, in {\it Relativity Today,
Proceedings of the 2nd Hungarian Relativity Workshop, Budapest, 1987},
edited by Z. Perj\'es (World Scientific, Singapore, 1988), p. 134.

\bibitem{Katz} J. Katz, Found. Phys. {\bf 33}, 223 (2003).

\bibitem{Demianski} M. Demia\'nski, {\it Relativistic Astrophysics}
({\it International Series in Natural Philosophy} vol~110) (Pergamon
Press/PWN-Polish Scientific Publishers, Oxfort/Warsaw, 1993).

\bibitem{NeugebauerAnn} G. Neugebauer, Ann. Phys. (Leipzig) {\bf 9},
SI-342 (2000) Spec. Issue.

\bibitem{Kleinwaechter} A. Kleinw\"achter, Ann. Phys. (Leipzig) {\bf 9},
SI-99 (2000) Spec. Issue.

\bibitem{ISM} G. Neugebauer and R. Meinel, J. Math. Phys. {\bf 44},
3407 (2003).

\end{thebibliography}
\end{document}